# Greedy-Mine: A Profitable Mining Attack Strategy in Bitcoin-NG

Junjie Hu[1], Zhe Jiang[2] and Chunxiang Xu[3]

[1] Department of Computer Science, University of Electronic Science and Technology of China, Chengdu 611731, China
hujj@std.uestc.edu.cn

[2] Department of Mathematical Sciences, University of Electronic Science and Technology of China, Chengdu 611731, China
zhejiang@std.uestc.edu.cn

[3] Department of Cyberspace Security, University of Electronic Science and Technology of China, Chengdu 611731, China
chxxu@uestc.edu.cn

**Abstract.** Bitcoin-NG is an extensible blockchain protocol based on the same trust model as Bitcoin. It divides each epoch into one Key-Block and multiple Micro-Blocks, effectively improving transaction processing capacity. Bitcoin-NG adopts a special incentive mechanism (i.e., the transaction fees in each epoch are split to the current and next leader) to maintain its security. However, there are some limitations to the existing incentive analysis of Bitcoin-NG in recent works. First, the incentive division method of Bitcoin-NG only includes some specific mining attack strategies of adversary, while ignoring more stubborn attack strategies. Second, once adversaries find a whale transaction, they will deviate from honest mining strategy to obtain extra reward. In this paper, we are committed to solving these two limitations. First, we propose a novel mining strategy named Greedy-Mine attack. Then, we formulate a Markov Decision Process (MDP) model to analyze the competition of honest miners and adversaries. Furthermore, we analysis the extra reward of adversaries and summarize the mining power proportion range required for malicious adversaries to launch Greedy-Mine to obtain extra returns. Finally, we make a backward-compatibility progressive modification to Bitcoin-NG protocol that would raise the threshold of propagation factor from 0 to 1. Meanwhile, we get the winning condition of adversaries when adopting Greedy-Mine, compared with honest mining. Simulation and experimental results indicate that Bitcoin-NG is not incentive compatible, which is vulnerable to Greedy-Mine attack.

# 1 Introduction

## 1.1 Related Work

In 2008, Nakamoto proposed the Bitcoin blockchain protocol, trying to achieve consensus under a permissionless setting [1]. Bitcoin blockchain, based on Proof of Work (PoW), effectively deters sybil attacks [2]. The blockchain can be seen as a decentralized ledger, which is composed of continuous blocks that follow certain rules and linked through specific cryptographic methods. In Bitcoin blockchain, the first block (which does not reference any other block) is called the genesis block. Each block is composed of block header and block body. The block header mainly includes hash of previous block, time stamp, etc. The block body includes complete transaction data. The successful applications of blockchain in the financial field [3, 4], the Internet of Things [5, 6, 7, 8], the network security field [9, 10], the public service field [11, 12], the digital copyright field [13, 14], and the insurance field [15], which have made blockchain technology widely concerned by all walks of life. In the process of continuous development of blockchain, its scalability problems are gradually emerging. Compared with the global payment system Visa with an average of 50000 TPS (Transactions Per Second), the current blockchain system, such as Bitcoin with an average of 7 TPS, ETH with an average of 20 TPS [16], and EOS with an average of 3000 TPS, is not enough to meet the needs of modern financial transactions. In Bitcoin, Nakamoto chooses a fairly secure system parameter, namely, the average block output time is 10 minutes and the block size is limited to 1MB. Relevant researches show that modifying the blockchain system parameters (such as increasing the block size limit or reducing the average block output time) can increase TPS to a certain extent, but will reduce the security level of the blockchain system [9, 17]. Therefore, redesigning the consensus protocol at the underlying blockchain has become a research hotspot in recent years.

The design of the new blockchain consensus protocol can be roughly divided into three categories: Block Classification, Parallel Chains, and Directed Acyclic Graph (DAG). In the area of block classification, FruitChain [18], Bitcoin-NG [19] divide blocks into two categories: the main blocks are responsible for choosing the longest chain of consensus protocol, and the micro blocks are responsible for packaging transactions, which can effectively improve the system throughput of blockchains. In parallel chains, OHIE [20] and Prism [21] can improve system throughput while ensuring system security. The design of Monoxide [22] is more complex. From the perspective of academic analysis, it can effectively improve throughput without certain security. And there is a trade-off between scalability and security in Monoxide. In a DAG-based design approach, Inclusive [23] only proposes basic design principles without detailed introduction to complement the protocol. In Spectre [24], transactions can be confirmed in seconds and throughput is increased by orders of magnitude over bitcoin. Phantom [25] uses a greedy algorithm to distinguish blocks mined by honest miners legally from blocks mined by malicious miners that deviate from the DAG mining protocol, and ultimately provides full order on the BlockDAG in a uniform manner by all honest nodes to meet the specific requirements for ledger timeline in smart contracts. In Conflux [26], it improves the performance of the blockchain through reasonable design and optimization of system, while ensuring the security of the blockchain. Conflux has improved the throughput of the blockchain at the consensus level and has reduced the waiting time of block confirmation. Among them, Bitcoin-NG blockchain has received extensive attention from blockchain practitioners.

Bitcoin-NG [19] is among the first and the most prominent PoW-based blockchains to approach the near-optimal throughput, which has the same trust model as Bitcoin. It divides blocks into two categories: Key-Blocks and Micro-Blocks. Key-Blocks are responsible for participating in consensus protocol, while Micro-Blocks are responsible for packaging transactions. Bitcoin-NG improves performance by separating consensus protocols and packaging transactions. More specifically, each Key-block is generated through the leader election process, and the corresponding leader will obtain a block reward, which is called mining process. Furthermore, the leader can package multiple Micro-blocks and receive transaction fees until the next key block is generated, which is called process of packaging transactions. More intuitively, Bitcoin-NG separates transaction serialization process from leader election process, which brings Bitcoin-NG to approach the near-optimal throughput, since Micro-blocks can be generated at a rate up to the network capacity. It is precisely for this reason that Bitcoin-NG has been applied to cryptocurrencies Waves [28] and Aeternity [29].

The idea of separation has inspired many novel blockchain protocols including ByzCoin [30], Hybrid consensus [31], Prism [21] and so on. Although these protocols can achieve lower latency or higher throughput than Bitcoin-NG, the design and analysis of their incentive mechanisms are still unclear. However, as the foundation of these protocols, Bitcoin-NG still has certain limitations in incentive analysis, which will be explained in detail in section 3.3.

In Bitcoin-NG, Eyal [19] proposes two possible malicious attacks (Transaction Inclusion Attack and Longest Chain Extension Attack), which derives the division proportion of transaction fees. Jiayuan Yin [27] proposes Modified Transaction Inclusion Attack, which reallocates transaction fees and improves the imperfection of original transaction inclusion attack. The above incentive analysis based on Bitcoin-NG only includes the limited mining attacks of adversaries, while ignoring more novel mining strategies. Besides, they do not consider the extreme cases that may occur in the blockchain, e.g., whale transaction. Once whale transactions are detected by adversaries, they will deviate from honest mining strategy to obtain extra reward (whale transactions are more profitable).

### 1.2 Our Contributions

To address the above issues, we first propose a novel mining strategy: greedy-mine, which can increase the reward of adversaries. Furthermore, we model greedy-mine strategy through Markov Decision Process (MDP) to analyze the competition of honest miners and adversaries. Finally, we model Markov Reward process and calculate the extra reward of adversaries, which indicates that Greedy-Mine is more profitable than honest mining strategy. Specifically, we have the following contributions:

We first represent the Bitcoin-NG incentive mechanism and visually redescribe the design principle of Bitcoin-NG, which greatly enhances the understanding of Bitcoin-NG's underlying design principle.

Second, we propose a novel mining strategy named Greedy-Mine and model Greedy-mine strategy through Markov Decision Process (MDP) to analyze the competition of honest miners and adversaries. We further calculate the extra reward of adversaries, which demonstrates that Bitcoin-NG mining is not incentive compatible.

Third, we summarize the mining power proportion range required for adversaries to launch Greedy-Mine to obtain excess returns. When the greedy pool has more than 18% of the system mining power, launching Greedy-Mine is more profitable than honest mining. Furthermore, miners with more mining power are more motivated to adopt Greedy-Mine.

Finally, we make a backward-compatibility progressive modification to Bitcoin-NG protocol that would raise the threshold of propagation factor from zero to 1. Meanwhile, we get the winning condition of adversaries when adopting Greedy-Mine and honest mining, respectively, which indicates Bitcoin-NG is vulnerable to Greedy-Mine attack.

## 2 Preliminaries

### 2.1 Overview of Bitcoin-NG

Bitcoin-NG is an extensible blockchain protocol based on the same trust model as Bitcoin. On the basis of Bitcoin blockchain, Bitcoin-NG improves the blockchain performance under the Nakamoto consensus by separating consensus protocols and packaging transactions. The time is divided into multiple epochs, and each epoch contains a leader (i.e., block in the main chain). The tenure of each leader is about 10 minutes, during which the transactions in the transaction pool will be packaged. Each leader can obtain corresponding block reward (coinbase reward) and transaction reward, which ensures that miners are willing to participate in the Bitcoin-NG.

### 2.2 Key-Block and Micro-Block

Bitcoin-NG divides blocks into two categories: Key-Blocks and Micro-Blocks. Key-Blocks are responsible for consensus agreements, meanwhile, Micro-Blocks are responsible for packaging transactions.

**Key-Blocks: Consensus Protocol.** Key-blocks are responsible for leader election, which ensures the security of consensus protocol. Similar to Bitcoin, the Key-Block contains reference to the previous block, current GMT time, coinbase transactions to pay out the reward, target value, nonce, and public keys for packaged micro-blocks. Miners must traverse nonces until the PoW Puzzle is successfully solved, which means the hash of Key-Block header smaller than the target. The miner who finds a valid key-block will set the coinbase transaction to output to his own account address, which is calculated through the hash of public key. The process of a miner trying a nonce can be seen as a Bernoulli trail. Multiple Bernoulli trails form a Bernoulli process. Therefore, the process of miners mining Key-Blocks is memoryless. Furthermore, Bitcoin-NG adjusts the difficulty of mining puzzle through changing the target value to maintain the average block generation rate, which ensures the security of Bitcoin-NG.

**Micro-Blocks: Packaging Transaction.** When a miner generates a valid Key-Block, he becomes the leader within the current epoch. Leader can package transactions to generate Micro-Blocks at a rate below the predefined maximum rate. The predefined maximum rate of the Micro-Blocks is deterministic and can be much higher than the average generation rate of the Key-Blocks, which increases the throughput of system. Therefore, leaders will generate Micro-Blocks with unpackaged transactions to obtain corresponding transaction fees. The Micro-Block header contains a reference to the previous block, the current GMT time, the hash of its account, and the signature of the Micro-Block header. Micro-Blocks are responsible for packaging transactions, which is critical to improve Bitcoin-NG's throughput. However, they have no contribution to consensus protocol.

### 2.3 Protocol of Bitcoin-NG

In Bitcoin-NG, the current local state of each node may be inconsistent due to frequent generation rate of Micro-Blocks, which brings forking. As shown in Fig. 1, when the Key-Block 1 is generated, the Micro-Blocks 1' and 2' may not have been received yet, which enables them to become orphan blocks. Meanwhile, transactions in these orphan blocks will not be executed. Therefore, users who detect the Micro-Blocks in the blockchain should wait for a period of network propagation until other Key-Blocks are generated on top of these Micro-Blocks (e.g., in Bitcoin, users need to wait for 6 blocks (approximately 60 minutes) to ensure that the blocks are executed).

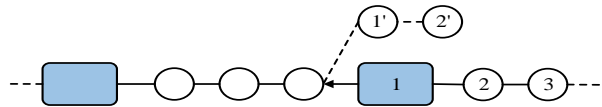

**Fig. 1.** Forking in Bitcoin-NG

To motivate miners to mine honestly and ensure the security of the system, leaders in each epoch will obtain two rewards: coinbase reward for generating Key-Blocks and transaction fees for generating Micro-Blocks. Meanwhile, the transaction fees should be shared by two adjacent leaders before and after the current epoch. Specifically, 40% of these transaction fees are earned by the leader of current epoch and 60% by subsequent leaders, as illustrated in Fig. 2 for details. The reason for choosing this distribution is explained in Section 3.

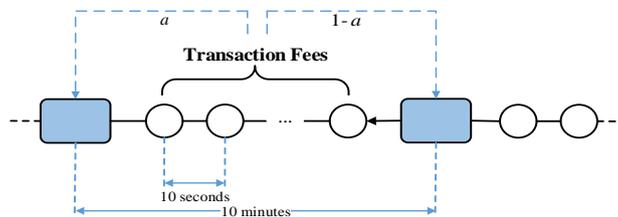

**Fig. 2.** Transaction incentives in Bitcoin-NG

## 3  Incentive Analysis of Bitcoin-NG

### 3.1  Original Incentive Analysis

Original incentive analysis of Bitcoin-NG contains two types of malicious attack strategies: Transaction Inclusion Attack and Longest Chain Extension Attack. We first define following parameters in Bitcoin-NG:

$\alpha$:     The total mining power of the adversary pool.

$\gamma$:     The ratio of other miners that choose to mine on the adversary's branch when forking competition occurs.

$r_{leader}$:     The ratio of transaction fees earned by the leader of current epoch.

**Transaction Inclusion Attack.** When adversaries generate a Key-Block and a series of Micro-Blocks with transactions, they may potentially increase their revenue of the transaction fees through selfish mining. To do so, adversaries first generate (reserve) Micro-Blocks with transactions, but do not publish them. Meanwhile, they try to mine on top of these unpublished Micro-Blocks, while other honest miners have to mine on published Key-Blocks. If adversaries find a subsequent Key-Block, they will publish it at once, which brings them 100% of the transaction fees (with probability $\alpha$). However, if other honest miners find a subsequent Key-Block and publish Micro-Blocks with these secret transactions, adversaries will try to mine on top of these Micro-Blocks, which brings them only $100\% - r_{leader}$ of the transaction fees (with probability $(1-\alpha) \cdot \alpha$). Fig. 3 shows the transaction inclusion attack in Bitcoin-NG. In order to urge all miners to adopt honest mining strategy, the revenue through transaction inclusion attach should be smaller than the revenue through honest mining. Therefore, we can get equation 1.

$$\overbrace{\alpha \cdot 100\%}^{\substack{sefish-mine\ succeed \\ (win\ 100\%)}} + \overbrace{(1-\alpha) \cdot \alpha \cdot (100\% - r_{leader})}^{\substack{mine\ succeed\ after\ tx \\ (win(100\%-r_{leader}))}} < \overbrace{r_{leader}}^{honest\ mine} \qquad (1)$$

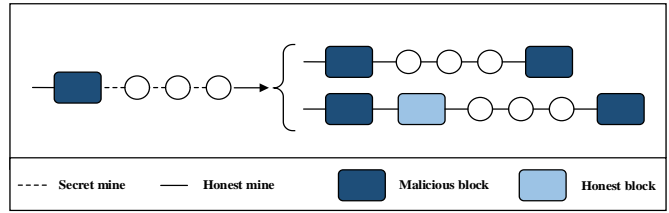

**Fig. 3.** Transaction inclusion attack in Bitcoin-NG

According to equation 1, we can get $r_{leader} > 1 - \frac{1-\alpha}{1+\alpha-\alpha^2}$. We assume that adversary owns the mining power $\alpha < \frac{1}{4}$, we can get $r_{leader} > 37\%$.

**Lonest Chain Extension Attack.** In order to improve revenue, the adversary can avoid Micro-Blocks, and directly mine on the previous Key-Block to generate a new valid Key-Block. Then he would generate Micro-Blocks with transactions. Once he finds a subsequent Key-Block, he will obtain $100\%$ of the transaction fees (with probability $\alpha^2$). Otherwise, he obtains $r_{leader}$ of the transaction fees (with probability $\alpha \cdot (1-\alpha)$). Fig. 4 shows the details of longest chain extension attack in Bitcoin-NG. The revenue that adversaries can obtain by longest chain extension attack must be smaller than the revenue obtained by honest mining. Therefore, we can derive equation 2.

$$\overbrace{\alpha^2 \cdot 100\%}^{Win\ 100\%} + \overbrace{\alpha \cdot (1-\alpha) \cdot r_{leader}}^{Win\ r_{leader}} < \overbrace{\alpha \cdot (1 - r_{leader})}^{Honest\ mine} \quad (2)$$

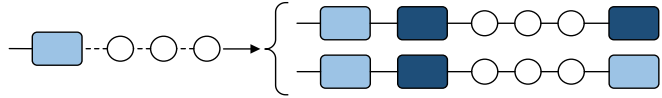

**Fig. 4.** Lonest chain extension attack in Bitcoin-NG

According to equation 2, we can get $r_{leader} < \frac{1-\alpha}{2-\alpha}$. Assume that adversary owns the mining power $\alpha < \frac{1}{4}$, we can get $r_{leader} < 43\%$.

### 3.2 Modified Incentive Analysis

**Modified Transaction Inclusion Attack.** Yin [27] improves Bitcoin-NG Transaction Inclusion Attack in relevant research, which modifies the transaction inclusion attack. More specifically, when adversaries find a valid Key-Blocks on top of these secret Micro-Blocks, they will publish these secret Micro-Blocks with transactions and the new valid Key-Block, which brings them $100\%$ of transaction fees (with probability $\alpha$). However, once honest miners find a valid Key-Block, they will publish it and Micro-Blocks with transactions. Meanwhile, adversaries will try to mine on top of these Micro-Blocks, which bring them $100\% - r_{leader}$ of transaction fees (with probability $(1-\alpha) \cdot \alpha$). Modified transaction inclusion attack in Bitcoin-NG is shown in Fig. 5. Therefore, we can derive equation 3.

According to equation 3, we can get $r_{leader} > \frac{\alpha}{1+\alpha}$. Assume that adversaries own the mining power $\alpha < \frac{1}{4}$, we can get $r_{leader} > 25\%$.

To sum up, we can get $\frac{\alpha}{1+\alpha} < r_{leader} < \frac{1-\alpha}{2-\alpha}$. Assume that adversaries own the mining power $\alpha < \frac{1}{4}$, we can get $20\% < r_{leader} < 43\%$. Therefore, the incentive parameter $r_{leader} = 40\%$ selected in Bitcoin-NG meets the security requirements.

$$\overbrace{\alpha \cdot 100\%}^{\substack{sefish-mine\ succeed \\ (win\ 100\%)}} + \overbrace{(1-\alpha) \cdot \alpha \cdot (100\% - r_{leader})}^{\substack{mine\ succeed\ after\ tx \\ (win(100\%-r_{leader}))}} < \overbrace{r_{leader} + \alpha \cdot (1 - r_{leader})}^{honest\ mine} \quad (3)$$

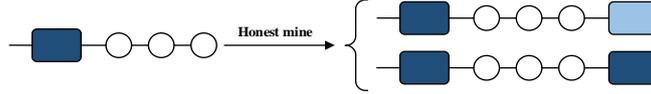

**Fig. 5.** Modified transaction inclusion attack in Bitcoin-NG

### 3.3 Defects of The Traditional Incentive Analysis

The above incentive analysis based on Bitcoin-NG only includes the limited mining strategies, while ignoring more stubborn mining strategies. For example, in the first stage, the adversary fails to package the whale transaction, but he may reverse the longest chain by generating new Key-Blocks, which could bring them more reward. On the basis of Bitcoin-NG's original incentive analysis above, we propose a novel mining strategy named Greedy-Mine and model it through Markov Decision Process (MDP) to analyze the competition of honest miners and adversaries (in section 4). We further calculate the extra reward of adversaries, which demonstrates that Bitcoin-NG mining is not incentive compatible (in section 5).

## 4 Markov Model of Greedy-Mine Strategy

### 4.1 Assumption

To simplify our analysis, we make some reasonable assumptions. Our assumptions are similar to those of other selfish mining attacks, such as selfish mining [32], stubborn mining [33], bribery semi-selfish mining and bribery stubborn mining [34].
1. We normalize the total mining power of the system to 1. The normalized mining power of adversary is a value greater than 0 but less than 1.
2. Miners are profit-driven. Honest miners can adopt the optimal mining strategy they consider to increase their profits, but will not launch mining attacks. This is reasonable because miners are honest but selfish. When the blockchain forks and the lengths of each branch are equal, miners could choose any branch.
3. There are no unintentional forks in the Bitcoin system. This assumption is rational because the probability of unintentional forks occurring in the Bitcoin system can be negligible, approximately 0.41% [35].
4. Block rewards can be ignored, compared with whale transaction. In our analysis, miner's rewards are expected as well as normalized.

## 4.2 Greedy-Mine Strategy

For the sake of simplicity and generality, we assume that mining power is divided into two categories: one is the minority mining pool following the Greedy-Mine strategy, and the other is the majority pool following the honest mining strategy. Furthermore, it is not significant whether honest miners are a single pool, a series of pools or individual miners.

The key intuition of Greedy-Mine strategy is that the greedy pool tries to compete with honest pool for the longest legal chain, which wastes the mining power of honest pool on the non-longest legal chain. Therefore, the greedy pool obtains all the transaction fees in each epoch. Therefore, adversaries have the motivation to exploit Greedy-Mine strategy to try to obtain excess returns.

When greedy pool finds a new Key-Block, he will publish it selectively, which makes greedy branch the longest legal chain and brings greedy pool all the transaction fees in each epoch. Generally, once a whale transaction occurs (whale transaction refers to the transaction involving very high transaction fees, and block rewards can be ignored, compared with whale transaction), greedy pool will try to generate Key-Blocks and Micro-Blocks with whale transactions, even if the whale transactions have been packaged into Micro-Blocks by other honest pool. Greedy pool will attempt to make their branch the longest legal chain, which wastes the mining power of honest pool. In this case, adversaries can obtain disproportionate reward under Greedy-Mine strategy.

With the above intuition, we propose Greedy-Mine strategy, which is driven by the mining events of greedy pool or honest pool. The decision of greedy pool is determined by the specific state of whale transactions. We divide $state_{tx}$ into three categories: $state_{tx} = 0$ refers that whale transaction has not been packaged, $state_{tx} = 1$ refers that whale transaction has been packaged by greedy pool and $state_{tx} = 2$ refers that whale transaction has been packaged by honest pool. The initialization of Greedy-Mine is described in the following algorithm 1.

| Algorithm 1. Initialization of Greedy-Mine |
| --- |
| 1: **On** Init |
| 2:     $state_{tx} = 0$ |
| 3:     $length(honest\ branch) = 0$ |
| 4:     $length(adversarial\ branch) = 0$ |
| 5:     Mine at the head of longest branch. |

**Fig. 6.** Initialization of Greedy-Mine

When greedy pool finds a Key-Block and whale transactions have not been packaged at this time, he will publish the Key-Block with whale transactions, which brings the $state_{tx}$ to 1 and adds one to the length of adversarial branch. If whale transactions have been packaged by greedy pool and the length of adversarial branch is not shorter than the length of honest branch, he will publish the Key-Block and add one to the length of greedy branch. If whale transactions have been packaged by honest pool and the length of honest branch is longer than adversarial branch, greedy pool will generate the Key-Block and add one to the length of the branch. The specific strategy of greedy pool is described in the following algorithm 2.

| Algorithm 2. Greedy-Mine strategy when greedy pool finds a Key-block |
|---|
| 1:  **On** Greedy pool finds a Key-block |
| 2:  $\Delta = length(adversarial\ branch) - length(honest\ branch)$ |
| 3:  **if** $state_{tx} == 0$ **then** |
| 4:      $state_{tx} = 1$ |
| 5:      $length(adversarial\ branch) + 1$ |
| 6:  **else if** $state_{tx} != 0$ **then** |
| 7:      **if** $\Delta < 0$ **then** |
| 8:          $length(adversarial\ branch) + 1$ |
| 9:      **else if** $\Delta\ !< 0$ **then** |
| 10:         Competition ends |
| 11: Mine at the head of greedy branch. |

**Fig. 7.** Greedy-Mine strategy when greedy pool finds a Key-block

When honest pool finds a Key-Block, he will continue to mine with honest strategy. If whale transactions have not been packaged, he will generate the Key-Block with whale transactions and set $state_{tx} = 1$. If whale transactions have been packaged, the strategy of honest pool is determined by the system state. If the length of honest branch is longer than greedy branch, honest pool will publish a Key-Block on honest branch and add one to the length of the honest branch. If the length of honest branch is equal to the length of the adversarial branch, the results of competition is determined by the choice of honest pool. Specifically, if honest pool appends honest branch, the length of honest branch adds one. Otherwise, the length of adversarial branch adds one. If the length of adversarial is longer than the length of honest branch, competition ends and adversarial pool gets all whale transaction fees. The specific strategy of honest pool is described in the following algorithm 3.

| Algorithm 3. Greedy-Mine strategy when honest pool finds a Key-block |
|---|
| 1:  **On** Honest pool finds a Key-block |
| 2:  $\Delta = length(adversarial\ branch) - length(honest\ branch)$ |
| 3:  **if** $state_{tx} == 0$ **then** |
| 4:      $state_{tx} = 2$ |
| 5:      $length(honest\ branch) + 1$ |
| 6:  **else if** $state_{tx} != 0$ **then** |
| 7:      **if** $\Delta < 0$ **then** |
| 8:          $length(honest\ branch) + 1$ |
| 9:      **else if** $\Delta > 0$ **then** |
| 10:         Competition ends |
| 11:     **else if** $\Delta == 0$ **then** |
| 12:         **if** honest pool appends honest branch **then** |
| 13:             $length(honest\ branch) + 1$ |
| 14:         **else if** honest pool appends adversarial branch **then** |

| | |
|---|---|
| 15: | $length(adversarial\ branch) + 1$ |
| 16: | Mine at the head of the longest branch |

**Fig. 8.** Greedy-Mine strategy when honest pool finds a Key-block

Under the Greedy-Mine strategy, greedy pool can obtain all whale transaction fees if he attacks successfully. On the contrary, nothing is gained.

### 4.3 State Transition Process

We model the state transition process of Greedy-Mine strategy in Fig. **9**. The state $s$ indicates that the whale transaction has not been packaged. The state $h_0$ indicates that the whale transaction is packaged by greedy pool and there are no Key-Blocks on top of it. The state $h_1$ is a termination state, which means that the whale transaction is packaged by greedy pool and another Key-Block on top of it is also generated by greedy pool. In this case, all the whale transaction fees are obtained by greedy pool. The states $h_{0k}(k \geq 1)$ indicate that the whale transaction is packaged by greedy pool and Key-Blocks on top of it are generated by honest pool. Meanwhile, the length of honest branch is $k+1$. The state $h_{10}$ indicates that the whale transaction is packaged by greedy pool, and the length of honest branch is equal to the greedy branch. The states $h_{1k}(k \geq 1)$ indicate that the whale transaction is packaged by greedy pool, and the length of honest branch is $k$ longer than greedy branch. The state $a_0$ indicates that the whale transaction is packaged by honest pool and no new Key-Blocks are on top of it. The state $a_1$ indicates that the whale transaction is packed by honest pool and an honest Key-Block is on top of it. The states $a_{0k}(k \geq 0)$ indicate that whale transactions are packaged by honest pool and honest Key-blocks are on top of it. Meanwhile, the length of honest branch is $k+2$. The state $a_2$ indicates that the whale transaction on the honest branch is packaged by the honest pool, and an honest block is on top of it. In this case, greedy pool may choose to launch Greedy-Mine to obtain more reward than honest mining. The states $a_{1k}(k \geq 0)$ indicate that the whale transaction on the honest branch is packaged by the honest pool, and an honest block is on top of it. Meanwhile, new honest Key-Blocks are generated on top of honest branch, and the length of honest branch is $k+1$ longer than that of greedy branch. The states $a_{2k}(k \geq 0)$ indicate that the whale transaction on the honest branch is packaged by the honest pool, and an honest block is on top of it. Meanwhile, new Key-Blocks are generated on top of both honest and greedy branch, and the difference between the length of honest branch and greedy branch is $k$. Next, we will discuss each state transition and probability in detail:

1. For state $s$: (1) when greedy pool finds a valid Key-Block, he will publish it with whale transactions, which brings the system to state $h_0$ (probability $\alpha$); (2) when honest pool finds a valid Key-Block, he will publish it with whale transactions, which brings the system to state $\alpha_0$ (probability $1 - \alpha$).
2. For state $h_0$: (1) when greedy pool finds a valid Key-Block, he will publish it on top of Micro-Block with whale transactions, which brings the system to state $h_1$ (probability $\alpha$); (2) when honest pool finds a valid Key-Block, he will publish it on top of Micro-Block with whale transactions, which brings the system to state $h_{00}$ (probability $1-\alpha$).
3. For state $h_1$: (1) when greedy pool finds a valid Key-Block, he will publish it on public chain, which brings the system to state $h_1$ (probability $\alpha$); (2) when honest pool finds a valid Key-Block, he will publish it on public chain, which brings the system to state $h_1$ (probability $1-\alpha$).
4. For states $h_{0k}(k \geq 1)$: (1) when greedy pool finds a valid Key-Block, he will publish it on top of Micro-Block with whale transactions and forking occurs, which brings the system to state $h_{1k}(k \geq 1)$ (probability $\alpha$); (2) when honest pool finds a valid Key-Block, he will publish it on public chain, which brings the system to state $h_{0(k+1)}(k \geq 1)$ (probability $1-\alpha$).
5. For state $h_{10}$: (1) when greedy pool finds a valid Key-Block, he will publish it on greedy branch, which brings the system to state $h_1$ (probability $\alpha$); (2) when honest pool finds a valid Key-Block, he will publish it on greedy branch or honest branch, which brings the system to state $h_1$ (probability $\gamma(1-\alpha)$) or state $h_{11}$ (probability $(1-\gamma)(1-\alpha)$).
6. For states $h_{1k}(k \geq 1)$: (1) when greedy pool finds a valid Key-Block, he will publish it on greedy branch, which brings the system to state $h_{1(k-1)}(k \geq 1)$ (probability $\alpha$); (2) when honest pool finds a valid Key-Block, he will publish it on honest branch, which brings the system to state $h_{1(k+1)}(k \geq 1)$ (probability $1-\alpha$).
7. For state $\alpha_0$: (1) when greedy pool finds a valid Key-Block, he will publish it with whale transactions on top of last published Key-Block, which brings the system to state $h_0$ (probability $\alpha$); (2) when honest pool finds a valid Key-Block, he will publish it on top of Micro-Block with whale transactions, which brings the system to state $\alpha_1$ (probability $1-\alpha$).
8. For state $\alpha_1$: (1) when greedy pool finds a valid Key-Block, he will publish it in front of whale transactions and forking competition occurs, which brings the system to state $\alpha_2$ (probability $\alpha$); (2) when honest pool finds a valid Key-Blocks, he will publish it on public chain, which brings the system to state $\alpha_{00}$ (probability $1-\alpha$).
9. For states $\alpha_{0k}(k \geq 0)$: (1) when greedy pool finds a valid Key-Block, he will publish it in front of whale transactions and forking competition occurs, which brings the system to state $\alpha_{1k}(k \geq 0)$ (probability $\alpha$); (2) when honest pool finds a valid Key-Block, he will publish it on public chain, which brings the system to state $\alpha_{0(k+1)}(k \geq 0)$ (probability $1-\alpha$).
10. For state $\alpha_2$: (1) when greedy pool finds a valid Key-Block, he will publish it on greedy branch, which brings the system to state $h_1$ (probability $\alpha$); (2) when honest pool finds a valid Key-Block, he will publish it on greedy branch or honest branch, which brings the system to state $h_{00}$ (probability $\gamma(1-\alpha)$) or state $\alpha_{10}$ (probability $(1-\gamma)(1-\alpha)$).
11. For states $\alpha_{1k}(k \geq 0)$: (1) when greedy pool finds a valid Key-Block, he will publish it on greedy branch, which brings the system to state $\alpha_{2k}(k \geq 0)$ (probability $\alpha$); (2) when honest pool finds a valid Key-Block, he will publish it on honest branch, which brings the system to state $\alpha_{1(k+1)}(k \geq 0)$ (probability $1-\alpha$).

12. For state $\alpha_{20}$: (1) when greedy pool finds a valid Key-Block, he will publish it on greedy branch, which brings the system to state $h_1$ (probability $\alpha$); (2) when honest pool finds a valid Key-Block, he will publish it on greedy branch or honest branch, which brings the system to state $h_1$ (probability $\gamma(1-\alpha)$) or state $\alpha_{21}$ (probability $(1-\gamma)(1-\alpha)$)
13. For states $\alpha_{2k}(k \geq 1)$: (1) when greedy pool finds a valid Key-Block, he will publish it on greedy branch, which brings the system to state $\alpha_{2(k-1)}(k \geq 1)$ (probability $\alpha$); (2) when honest pool finds a valid Key-Block, he will publish it on honest branch, which brings the system to state $\alpha_{2(k+1)}(k \geq 1)$ (probability $1-\alpha$).

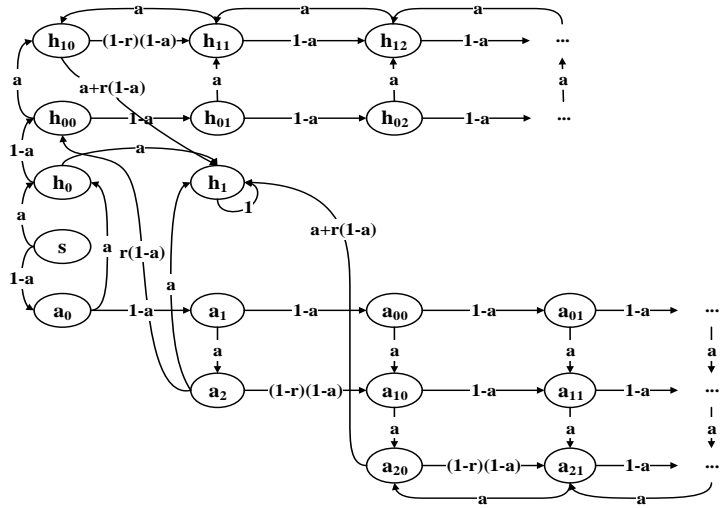

**Fig. 9.** State transition process of Greedy-Mine

### 4.4 Equation of State Probability

According to the state transition process in Fig. 9, we can derive the equation 4.

$$\begin{cases} p_s = 1 \\ p_{a_0} = (1-\alpha)p_s \\ p_{h_0} = \alpha + \alpha \cdot p_{a_0} \\ p_{h_{00}} = (1-\alpha)p_{h_0} + \gamma(1-a)p_{a_2} \\ p_{h_{10}} = \alpha(p_{h_{00}} + p_{h_{11}}) \\ p_{a_1} = (1-\alpha)p_{a_0} \\ p_{a_2} = \alpha \cdot p_{a_1} \\ p_{a_{00}} = (1-\alpha)p_{a_1} \\ p_{a_{10}} = (1-\gamma)(1-\alpha)p_{a_2} + \alpha \cdot p_{a_{00}} \\ p_{a_{20}} = \alpha \cdot p_{a_{10}} + \alpha \cdot p_{a_{21}} \end{cases} \quad (4)$$

We divide the state probability transition process into two parts. Since the termination state probability $p_{h_1}$ is determined by the probabilities of state probability $p_{a_{20}}$ and $p_{h_{10}}$, we analyze state probability $p_{a_{20}}$ and $p_{h_{10}}$ respectively. Furthermore, we can derive the state probability $p_{a_{20}}$ through the state probability $p_{a_{10}}$ (as shown in equation 5). Similarly, the state probability $p_{h_{10}}$ can be expressed by the state probability $p_{h_{00}}$ (as shown in equation 6).

$$p_{a_{20}} = \alpha \cdot p_{a_{10}} \cdot \sum_{k=0}^{\infty} \alpha^k (1-\alpha)^k \quad (5)$$

$$p_{h_{10}} = \alpha \cdot p_{h_{00}} \cdot \sum_{k=0}^{\infty} \alpha^k (1-\alpha)^k \quad (6)$$

According to the equation 4-6, we can derive the following equations:

$$\begin{cases} p_{h_0} = \alpha(2-\alpha) \\ p_{a_2} = \alpha(1-\alpha)^2 \\ p_{h_{10}} = \dfrac{\alpha^2(1-\alpha)(2-\alpha) + \gamma\alpha^2(1-\alpha)^3}{1-\alpha(1-\alpha)} \\ p_{a_{20}} = \dfrac{(2-\gamma)\alpha^2(1-\alpha)^3}{1-\alpha(1-\alpha)} \end{cases} \quad (7)$$

## 5 Revenue Analysis

### 5.1 Revenue Analysis of Honest Mine

(1) When adversary pool finds a Key-Block with whale transactions and then finds another Key-Block on top of it, he obtains 100% of whale transaction fees (probability $\alpha^2$).

(2) When adversary pool finds a Key-Block with whale transactions and then honest pool finds another Key-Block on top of it, adversary pool obtains $r_{leader}$ of whale transaction fees (probability $\alpha(1-\alpha)$).

(3) When honest pool finds a Key-Block with whale transactions and then adversary pool finds another Key-Block on top of it, adversary pool obtains $100\% - r_{leader}$ of whale transaction fees (probability $(1-\alpha)\alpha$).

According to the revenue analysis of honest mining, we calculate the revenue expectation that the honest pool with mining power of $a$ can obtain, as shown in equation 8.

$$r_{Honest-Mine} = \overbrace{\alpha^2 \cdot 100\%}^{Case(1)} + \overbrace{\alpha(1-\alpha) \cdot r_{leader}}^{Case(2)} + \overbrace{(1-\alpha)\alpha(100\% - r_{leader})}^{Case(3)} \quad (8)$$

### 5.2 Revenue Analysis of Greedy-Mine

(4) When state $h_0$ transitions to termination state $h_1$, greedy pool can obtain revenue of $\alpha \cdot p_{h_0} \cdot 100\%$.

(5) When state $h_{10}$ transitions to termination state $h_1$, greedy pool can obtain revenue of $(\alpha + \gamma(1-\alpha)) \cdot p_{h_{10}} \cdot 100\%$.

(6) When state $\alpha_2$ transitions to termination state $h_1$, greedy pool can obtain revenue of $\alpha \cdot p_{a_2} \cdot 100\%$.

(7) When state $\alpha_{20}$ transitions to termination state $h_1$, greedy pool can obtain revenue of $(\alpha + \gamma(1-\alpha)) \cdot p_{a_{20}} \cdot 100\%$.

According to the revenue analysis of Greedy-Mine, we calculate the revenue expectation that greedy pool with mining power $\alpha$ can obtain, as shown in equation 9.

$$r_{Greedy-Mine} = \overbrace{a \cdot p_{h_0} \cdot 100\%}^{Case(4)} + \overbrace{(\alpha + \gamma(1-\alpha))p_{h_{10}} \cdot 100\%}^{Case(5)} \\ + \overbrace{(\alpha \cdot p_{a_2} \cdot 100\%}^{Case(6)} + \overbrace{(\alpha + \gamma(1-\alpha))p_{a_{20}} \cdot 100\%}^{Case(7)} \quad (9)$$

### 5.3 Simulation and Experimental Result

We next present a systematic evaluation of the revenue of the adversary exploiting Greedy-Mine strategy. Furthermore, we evaluate the minimum mining power threshold that greedy pool is willing to exploit Greedy-Mine strategy to obtain disproportionate reward. Fig. 10 shows the revenue that adversary with different mining power can obtain by launching Greedy-Mine strategies under different propagation factor parameters, compared with the honest mining protocol in Bitcoin-NG.

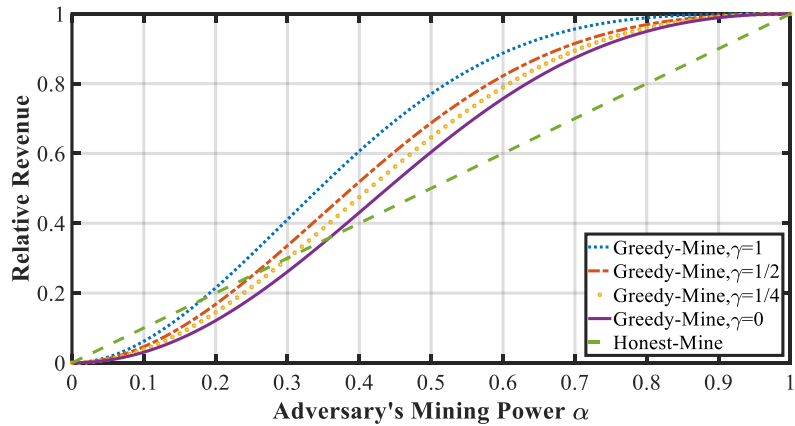

**Fig. 10.** The revenue of greedy pool that adopts Greedy-Mine strategy for different propagation factors $\gamma$, compared with honest mining in Bitcoin-NG.

Fig. 10 indicates that the simulation results are consistent with the theoretical analysis, both of which show that greedy pool with higher minting power will obtain higher revenue by adopting Greedy-Mine strategies. Moreover, it demonstrates that the Bitcoin-NG mining is not incentive compatible even in the presence of honest pool majority. More specifically, we set the propagation factor to four cases ($\gamma = 0, 0.25, 0.5, 1$). The experimental results show that when $\gamma = 1$, the minimum mining power owned by greedy pool to launch Greedy-Mine is $\alpha = 0.18$. Furthermore, once the greedy pool possesses more than 18% of mining power, adopting Greedy-Mine is always the optimal mining strategy, which could bring him more reward, compared with honest mining. Fig. 11 shows the minimum mining power owned by the adversary that could obtain disproportionate revenue under different propagation factor parameters $\gamma$ when adopting Greedy-Mine, compared with honest mining. Solid line represents no extra reward, which means that no matter whether Greedy-Mine or honest mining is adopted, the reward of adversary is the same. The right side of solid line indicates that adopting Greedy-Mine is optimal strategy. Furthermore, the winning area of Greedy-Mine is larger than honest mining, which is consistent with our theoretical analysis.

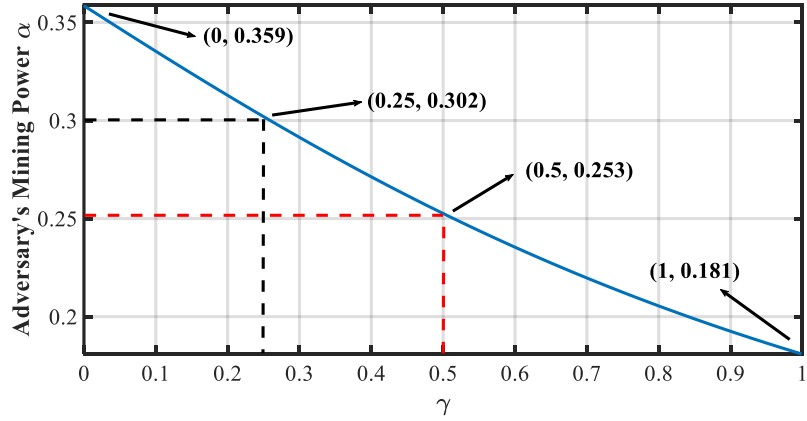

**Fig. 11.** The minimum mining power owned by the adversary that could obtain disproportionate revenue under different propagation factor parameters $\gamma$ when adopting Greedy-Mine, compared with honest mining.

When the propagation factor $\gamma$ is larger, honest miners are more likely to contribute to the greedy branch. The minimum mining power owned by the adversary to exploit the Greedy-Mine strategy to obtain extra revenue is 0.18, compared with honest mining in Bitcoin-NG, which indicates Bitcoin-NG mining is not incentive compatible.

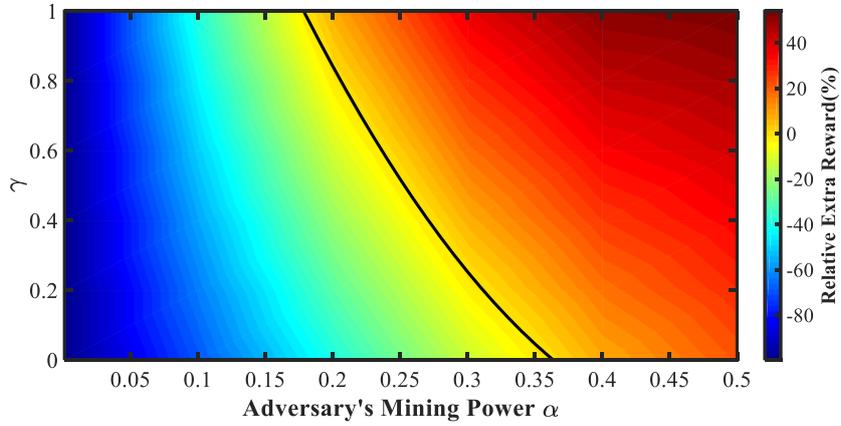

**Fig. 12.** The minimum mining power owned by the adversary that could obtain disproportionate revenue under different propagation factor parameters $\gamma$ when adopting Greedy-Mine, compared with honest mining.

Considering the adversary with mining power $\alpha$ in a mining game under Greedy-Mine and honest mining. We define the winning condition for adversary is obtaining a higher reward than honest mining strategy. To provide a more detailed description, we

define relative extra reward (RER) to show the performance of Greedy-Mine, which can be expressed as follows:

$$RER^{\tau_1,\tau_2} = \frac{R^{\tau_1} - R^{\tau_2}}{R^{\tau_2}} \times 100\% \qquad (10)$$

Where $\tau_1$, $\tau_2$ indicate different mining strategy (i.e., Greedy-Mine or honest mining), and $R^{\tau_1}$ represents the reward of adversary when adopting $\tau_1$ mining strategy.

We show the relative extra reward and winning condition of adversary in Fig. 12. More specifically, the right side of solid line is the winning condition of adversary when adopting Greedy-Mine. When $\alpha$ and $\gamma$ are relatively large, adversary could obtain higher relative extra reward under Greedy-Mine. The reason is that the more miners choose to mine on greedy branch in forking competition, the higher probability of adversary winning. Meanwhile, comparing with honest mining, miners with lager mining power have an advantage in adopting Greedy-Mine. Therefore, Miners with relatively large mining power have the motivation to use Greedy-Mine. The experimental result of adversary's relative extra reward with different mining power $\alpha$ under different propagation factor parameters $\gamma$ is given in Table 1. The experimental result indicates that the adversary with more mining power has motivation to adopt Greedy-Mine strategy, regardless of propagation factor parameters $\gamma$. Moreover, with the increasing of $\gamma$, the adversary can obtain more relative extra reward than honest mining strategy, regardless of adversary's mining power $\alpha$, which is consistent with our theoretical analysis.

**Table 1.** Adversary's relative extra reward with different mining power $\alpha$ under different propagation factor parameters $\gamma$.

|  | $\alpha = 0.1$ | $\alpha = 0.2$ | $\alpha = 0.3$ | $\alpha = 0.4$ | $\alpha = 0.5$ |
| --- | --- | --- | --- | --- | --- |
| $\gamma = 0$ | -69.4187 | -39.4667 | -12.9279 | 7.7053 | 20.8333 |
| $\gamma = 0.2$ | -63.1523 | -30.0900 | -2.9542 | 16.4968 | 27.5000 |
| $\gamma = 0.4$ | -56.8859 | -20.6933 | 7.0195 | 25.2884 | 34.1667 |
| $\gamma = 0.6$ | -50.6196 | -11.3067 | 16.9932 | 34.0800 | 40.8333 |
| $\gamma = 0.8$ | -44.3532 | -1.9200 | 26.9668 | 42.8716 | 47.5000 |
| $\gamma = 1$ | -38.0868 | 7.4667 | 36.9405 | 51.6632 | 54.1667 |

## 6 Conclusion

Although Bitcoin-NG is scalable, it is vulnerable to the novel mining attacks (e.g., Greedy-Mine). In our work, we present a novel mining attack named Greedy-Mine and demonstrate that in Pow-based blockchain system such as Bitcoin-NG, Greedy-Mine strategy can bring adversaries more relative extra reward, compared with honest mining. Furthermore, comparing with honest mining, once the adversary possesses more than 18% of mining power, adopting Greedy-Mine is always the optimal mining strategy, which could bring him more reward. Finally, we get the winning condition of adversaries when adopting Greedy-Mine and honest mining, respectively, which indicates Bitcoin-NG is vulnerable to Greedy-Mine attack. Finally and hopefully, the Greedy-Mine strategy that we propose in this paper can serve as a crucial reference for future researches of blockchain.